\documentclass[12pt]{article}
\usepackage[]{graphicx}
\usepackage{authblk}
\DeclareGraphicsExtensions{.pdf,.jpg}

\usepackage{amsfonts,amsmath,amsthm,amssymb}
\usepackage{mathrsfs}
\usepackage{dsfont}
\usepackage{appendix}
\usepackage[english]{babel}
\usepackage[onehalfspacing]{setspace}
\usepackage{hyperref}
\usepackage{pdfsync}
\usepackage{color}

\newcommand{\beq}{\begin{equation}}
\newcommand{\eeq}{\end{equation}}
\newcommand{\beqr}{\begin{eqnarray}}
\newcommand{\eeqr}{\end{eqnarray}}
\author[1,2]{Sujoy K. Modak\thanks{smodak@ucol.mx}}
\affil[1]{\small{Facultad de Ciencias - CUICBAS, \\Universidad de Colima, CP 28045, Colima, Mexico}} 
\affil[2]{\small{California State University, Fresno, CA 93740-8031, USA}}
\title{{Does gravity cause disentanglement?}}
\begin{document}

\maketitle

\begin{abstract}
  Inspired by our recent works on information paradox in black holes, which exploit various foundational intricacies of quantum mechanics,  here we propose  a novel connection between the spacetime geometry and quantum entanglement of matter fields ``living'' in that geometry. We highlight, as a natural consequence of those studies, that gravitational field might have a natural tendency of reducing entanglement between two quantum states. 
\end{abstract}

{\it Allowing the loss of information in black holes:} 

Recently we found an acceptable path  \cite{BH, BH2, BH3} to account for the loss of information during a black hole evaporation \cite{bhp}. These proposals have overcome some important initial challenges and, made an important step forward by uniting this ``black hole problem'' with the so called ``measurement problem'' of quantum theory \cite{mp, WB, Schlosshauer:2003zy, penrose}. Here, we push those ideas further to advocate a possible connection between the entanglement of quantum states  and the curvature of spacetime they ``live on''. This possibility arises as a logical extension of our works  \cite{BH, BH2, BH3}, which, on the other hand, are based on the realisation that,  quantum dynamics of matter fields is never unitary by 100\%, as highlighted over and over again by Collapse Models \cite{revcol}. Allowing the collapse process, as a physical counterpart, ramifies various unanswered questions of the Copenhagen interpretation of quantum mechanics. These models, first built in a non-relativistic framework, united the continuous (and unitary evolution) and collapse (and stochastic reduction) dynamics of wavefunctions by a new unified evolution law. Recently, relativistic versions were also built along these lines \cite{relcol, relcol2}. Contrary to Copenhagen interpretation, collapse models always allow, a slight possibility, for the wavefunction to spontaneously collapse. The idea, behind allowing collapse for all practical purposes, is  not only to explain the laboratory experiments (for which the collapse time is very small) but also various natural processes out there (for which rate of collapse varies) where no observer is making a measurement. In Copenhagen interpretation, collapse happens only under the influence of a lab based measurement, and not when the system is not measured. Therefore, naturally, allowing a nonzero collapse in the absence of laboratory measurement, needs a physical explanation.  Indeed, there is an increasingly growing consensus, that originated from the works of Penrose and D\'iosi, that this spontaneous collapse of the wavefunction, even in the absence of measurement, is triggered by the gravitational disturbance \cite{GSR}. The point is that,  as a theory, quantum mechanics or its relativistic counterpart, is built using the notion of space and time, and in a realistic universe the spacetime is never flat since it is filled with matter and the quantum system cannot be shielded from this gravitational influence. So practically speaking, a quantum system is never closed from gravitational influence, and this very influence causes the tendency to collapse the wavefunction. More arguments in these lines can be found in \cite{GSR, revcol}. 

Although, these models are yet to be verified experimentally \cite{GSR, test}, one immediately sees an advantage of the broad idea in building a case \cite{elias} to say, that, indeed it is possible to lose information in a black hole evaporation. This is because the initial ``in'' vacuum state corresponding the initial state of matter field, during evaporation process, inevitably passes through an ultra large gravitating field (inside black holes), and undergoes a spontaneous collapse due to gravity. Since this collapse is {\it stochastic}, such as required by the collapse models to reproduce the Born rule, there is a violation of unitarity during collapse. To address information loss,  the rest is, to carefully build models with finer details and, to really show this is the case \cite{BH, BH2, BH3}. Barring the detailed technicality, it is quite simple and elegant to demonstrate a gravitationally induced generalized evolution of the ``in'' vacuum state, as well as, the evolution of the initial (pure) density matrix,  during the black hole evaporation process.

\noindent
Consider the time evolution of the initial ``in'' vacuum $|\Psi_i \rangle = |in\rangle$, which can be expressed as a linear combination of all excited states, in the ``out'' region, as
\beq
| \psi_{i} \rangle = N\sum_{F}e^{-\frac{\beta E_F}{2}}|F\rangle^{int}\otimes|F\rangle^{ext}.
\label{in}
\eeq
Here, $N$ is a normalisation constant, $\beta$ is the inverse Hawking temperature, the states $|F\rangle^{int/ext}$ are the {\it entangled} particle excited states forming Fock bases for the interior (to the event horizon) and exterior Hilbert spaces, in a sense that they exist pairwise and, the particle content of the exterior state is {\it entangled} with the anti-particle content of the interior state.  If we now allow the possibility, that alongside the Schrodinger dynamics, there is a gravitational collapse of quantum superpositions, we shall need a unified framework for both processes. The unified evolution can be proposed by the so called Continuous-Spontaneous-Localisation theory \cite{csl}, which in our case is given by
\beqr \label{CSL-QFT3}
| \psi (t) \rangle &=&{\cal T}e^{-\int_{0}^{t}dt'\big[\frac{1}{4\lambda} \sum_{nj}
[w_{nj}{} (t')-2\lambda\hat N_{nj}]^{2}\big]} | \psi_{i} \rangle.
\eeqr
\begin{equation}
\rho(\tau)=\mathcal{T}e^{-\int_{\tau_{0}}^{\tau}d\tau'\frac{\lambda(\tau')}{2}\sum_{n,j}[N_{n,j}^L-N_{n,j}^R]^2}\rho(\tau_{0}).\label{dm}
\end{equation}
where, ${\cal T}$ is the time ordering operator, $N_{n,j} = N_{n,j}^{int} \otimes \mathds{1}^{ext}$ is the number operator, made up from the direct product of the number operator of the interior (to event horizon)  basis and identity operator of the exterior basis, with quantum numbers $n,j$ ($n$ being the number of particle (or anti-particle) excitation and $j$ corresponds to discrete energy). Collapse happens to the eigenbasis of this operator. The collapse parameter $\lambda$ determines the rate of collapse. The function $w_{nj}$ is a classical stochastic function of time of white noise type distribution, determining which one among the eigenvectors will be realized, post collapse. Even if we start with identical copies of the same initial state, each of them, will evolve to different final states, post-collapse, simply because $w_{nj}$ are stochastically  chosen. Before demonstrating the CSL evolution, it is important to comment on two things - (i) the above equation is valid for the interaction picture and therefore the free Hamiltonian does not appear in \eqref{CSL-QFT3} and \eqref{dm}, and (ii) the collapse parameter is hypothized to be a monotonously increasing function of the Weyl curvature square $\lambda = \lambda(W_{abcd}W^{abcd})$ (by following Penrose Weyl curvature hypothesis). As the singularity is approached Weyl curvature scalar diverges, making $\lambda$ to diverge  and, this  intensifies collapse process enormously by breaking superpositions. 

A formal evolution of the initial state requires a foliation of the black hole spacetime in terms of Cauchy slices. We made an explicit slicing of the CGHS black hole spacetime \cite{BH}, RST spacetime \cite{BH2} to demonstrate the CSL dynamics. Further, we also proved the foliation independence in \cite{BH3}. The key point of the evolution is the following: we first note that the ``in'' vacuum state can be expressed, in the joint internal and external basis to the black hole, as in \eqref{in}. Then while we evolve it through the Cauchy slices, the interior part evolves towards region of high curvature, therefore the superposition of particle excited states in the interior basis, are the ones and {\it only ones}, tend to collapse to one of the eigenstates of the number operator $N_{n,j}$, due to CSL dynamics. By definition, this operator does not do anything {\it directly} to the particle states in the exterior region. Nevertheless, note that, the exterior states $|F_{nj}\rangle^{ext}$ (reaching asymptotic observer) also gets affected because these states are entangled with $|F_{nj}\rangle^{int}$. Basically, the exterior states also suffers a collapse, but not directly by gravity, rather due to the fact that their entangled counterparts are collapsed due to gravity. Therefore, if the particle excited state in the interior basis collapses to a Fock state, say  $|F_0\rangle^{int}$, where  $F_0 = \lbrace F_0^{nj} \rbrace $ is a  complete   set of  the occupation numbers  in each mode, then the exterior state collapses,  due to entanglement, to $|F_0\rangle^{ext}$. This yields the post-collapse final state as  $| \psi_{CSL} \rangle = N C_{F_0}|F_0\rangle^{int}\otimes|F_0\rangle^{ext}$. Once they are collapsed the entanglement between them is broken and all of this due to gravity! Note that, however, that the final state is pure, although due to stochasticity in $w_{nj}(t)$, it will be {\it undetermined} making the loss of predictability which is usually associated with the breaking of unitary evolution. Similarly, one can show quite easily \cite{BH} the density matrix \eqref{dm}, after collapse is mixed with a thermal weight, given by
\beqr
\rho_{final} = \rho^{thermal}
\eeqr
Therefore, we see, although the initial state remains pure after evaporation, it becomes {\it undetermined}. As a result, at the ensemble level, the density matrix  becomes mixed. This evolution is highly non-unitary, however, it is no threat to the established physical laws, rather a hint of novel dialogue between quantum dynamics under gravitational influence.

{\it A general picture - gravity and entanglement:}
What does the loss of information teach us about the quantum entanglement in a curved background? We seek an answer to this question here.

To understand that, once again, we notice the fact that, in our proposal the ``in'' vacuum state collapses into a stochastically chosen particle excited state of the interior Hilbert space, due to gravitational influence and that,  by entanglement, ultimately the superposition in the exterior Hilbert space is collapsed. This phenomena, of course, has nothing to do with the existence of the event horizon, rather, only depends on local curvature. Therefore, it should take place, even in the absence of a horizon, whenever a quantum state is evolved through a region of non-zero spacetime curvature. The difference for the latter, from the black hole case, will be an incomplete (weak) reduction of quantum state as compared to an almost complete (strong) reduction for black holes. The point of discussion here is to explain the implication of this for a pair of entangled states.

Speaking, a bit loosely, now in the language of entangled particles, consider a pair creation where both particles are entangled with each other, from their birth and, they travel distinct patches of the spacetime.  During their travel, let us assume that particle 1 reaches a region of spacetime of comparatively large curvature while particle 2  always sees a flat portion of the spacetime. In this case, particle 1 experiences a gravitational induced collapse, which may also be incomplete, and by entanglement, the other particle's state also gets affected. By incomplete collapse we mean that the purity of the quantum state is increased and as a result channel of entanglement become noisy so that the status of maximal entanglement gets affected. With a complete collapse, this two particle state will be represented by pure states (a direct product of individual particle states without superpositions). Now, the natural implication of this is to realise, that the entangled states which were entangled in past may not stay like that forever in future, even without someone measuring this. Gravitational field is enough to disentangle the pairs!

For an example, we can think of our own galaxy - the AGN black hole that at the centre of our galaxy which is about 26000 light years away from us. Imagine that, an entangled pair of particles, say photons, one of which goes towards the centre of the black hole and the other goes elsewhere. If the photons were created from the same source, initially they had polarisations, left and right handed, entangled. If we think a time much earlier than 26000 years, and the situation in which one of the photons travelled toward the centre of our galaxy and eventually entering the black hole, by today the quantum state of the photon might have already collapsed with one of the polarisations, and that in turn, could have made, by the property of entanglement, the polarisation state of the other photon, travelling elsewhere, also to take a specific orientation. This way a  natural purification for the state of a  two entangled photons might take place in nature.  Similar discussion applies to other entangled physical parameters, such as, position, momentum or spin.

{\it Discussion:} Understanding entanglement is a subject of intense research and we believe the lesson from black hole information paradox sketches an important starting point for this, in a curved space. This, at the same time, brings a possibility to test of our argument to resolve black hole information paradox, by measuring the disentanglement between particle pairs in curved space, as well as, an opportunity to test gravity induced collapse which is very important to resolve the measurement problem of quantum theory. Of course, it needs a great deal of future work and, we shall also need to do so with the available technological capacity. Nevertheless, testing the survival of entanglement at larger distances is an evolving topic for experimentalists. The new world record is about 1400 Kms \cite{teleport}. We wonder, if that can be extended to astrophysical scale to test our proposal, and, we do not know any compelling reason to dismiss such a possibility in future.

%%%%%%%%%%%%%%%%%%%%%%%%%%%%%%%%%%%%%%%%%%

\end{document}